\documentclass[a4paper]{article}

\usepackage{INTERSPEECH2021}
\title{Adding Connectionist Temporal Summarization into Conformer to Improve
Its Decoder Efficiency For Speech Recognition}
\name{Nick J.C. Wang, Zongfeng Quan, Shaojun Wang, Jing Xiao}

\address{
  Ping An Technology}
\email{}

\begin{document}

\maketitle
\begin{abstract}
The Conformer model is an excellent architecture for speech recognition modeling that
effectively utilizes the hybrid losses of connectionist temporal classification
(CTC) and attention to train model parameters. To improve the decoding efficiency of
Conformer, we propose a novel connectionist temporal
summarization (CTS) method that reduces the number of frames required for the
attention decoder fed from the acoustic sequences generated by the encoder,
thus reducing operations. 
However, to achieve such decoding improvements, 
we must fine-tune model parameters, 
as cross-attention observations are changed 
and thus require corresponding refinements. 
Our final experiments show that, 
with a beamwidth of 4, 
the LibriSpeech's decoding budget can be reduced 
by up to 20\% and 
for FluentSpeech data it can be reduced by 11\%, 
without losing ASR accuracy.
An improvement in accuracy is even found for the LibriSpeech "test-other" set. 
The word error rate (WER) is reduced by 6\% relative at the beam width of 1 and by 3\% relative at the beam width of 4.
\end{abstract}

\section{Introduction}

End-to-end models have prevailed in ASR in recent years. 
These directly transform speech feature sequences to label sequences using
a single neural network,
which smoothly integrates the acoustic model, pronunciation model, and
language model into one network. Approaches of this type typically fall into 
one of three categories.
CTC began the trend with its special dynamic programming design by merging
alignments with the same label output sequence~\cite{Graves2006,Graves2014}.  
It considers more decoding paths in the same training data than 
cross-entropy (CE) training, in which different paths may represent
different pronunciations of the same label sequence so as to enhance the
robustness of the model. 
However, CTC still assumes conditional independence and only poorly utilizes
the language knowledge of word association. 
RNN-Transducer follows CTC's dynamic programming to compute the marginal
probability, 
but adds a label predictor network, conditions on previous output labels, 
and decides the final output labels jointly with both the original acoustic
encoder and the additional linguistic predictor~\cite{Graves2013SpeechRW}.
In this way, it deeply augments the language model into an end-to-end neural
model and predicts the next label conditioned on the historical output
labels.  
These models are all based on RNN and a monotonic alignment assumption.  
The attention-based encoder-decoder architecture includes the previous output
labels as decoding conditions as does RNN-Transducer, but also utilizes
a special attention scheme that allows the decoder in each step to attend the
encoding sequence in full context~\cite{Chan2016ListenAA}.
This attention broadly expands the context to the sentence level.
Transformer pushes the use of attention from cross attention to 
self-attention~\cite{Vaswani2017,Dong2018SpeechTransformerAN,Vila2018EndtoEndST},
and Conformer adds convolution network layers into Transformer and improves its
robustness and accuracy~\cite{Gulati2020ConformerCT}. 

The attention-based approach was first invented for translation tasks to
great success.
Sentence-level attention allows it to translate words from the source language
to the target language without word-order constraints:
the next word can be predicted by conditioning on any words in the sentence.
The multi-head attention mechanism further enhances its dependency modeling.
Later CTC/attention multi-target learning (MTL) approaches have proven 
helpful in training attention-based models, by using CTC loss to constrain the
encoder~\cite{Watanabe2017, Miao2019}. 

In this paper, we seek to further improve the computational efficiency of
Conformer.  
We propose the CTS segmental representation. For the Conformer model, a mask is used to
easily augment CTS into Conformer's modeling.
We describe the algorithm in the next section along with fine-tuning
approaches to ensure its accuracy. 

For speech modeling, compact speech
representations have been developed~\cite{YuAnChung2016,YuAnChung2018,YuhsuanWang2018}, including
segmental representation~\cite{YuhsuanWang2018}. 
However, these involve training an independent autoencoder network to produce
speech representations, which may not be optimized for ASR purposes. 
For example, Wang et~al.\ generate segmental boundaries via special segmentation 
gates~\cite{YuhsuanWang2018}.
In contrast, in our approach, as the segmental representations are calculated from the
CTC module trained for ASR, there are no complicated autoencoder networks to train. 
Also, audio word embeddings can be aligned with text word embeddings to broaden the 
training data to enhance the model~\cite{YichenChen2018,YinghuiHuang2020}.

\section{CTS-Conformer}

\begin{figure}[t]
  \centering
  \includegraphics[width=\linewidth]{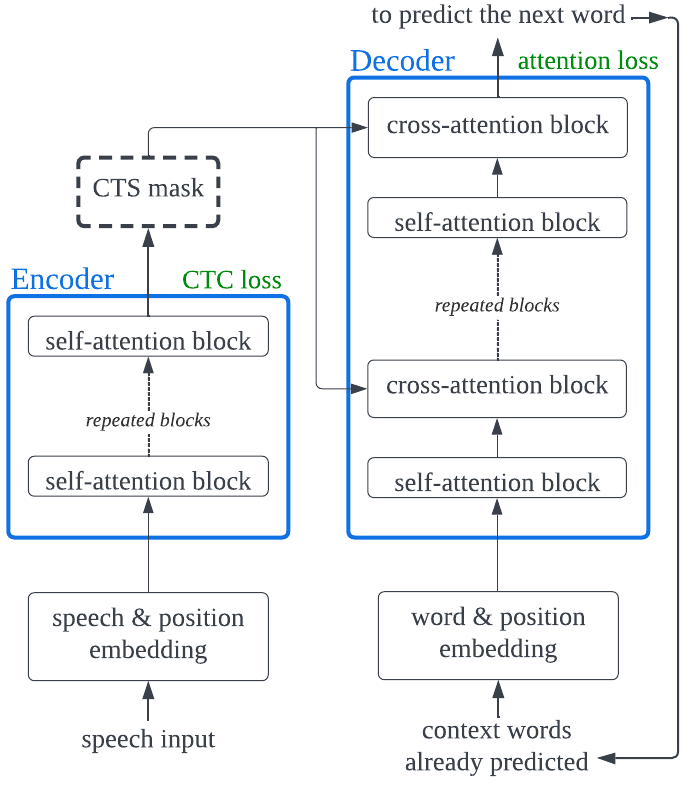}
  \caption{CTS-Conformer structure}
  \label{fig_cts_conformer}
\end{figure}

The CTS-Conformer adds a single CTS component into the Conformer model to form the
following structure, as shown in Fig.~\ref{fig_cts_conformer}:
\begin{itemize}
\item Self-attention encoder
\item CTC loss for encoder
\item CTS mask
\item Cross-attention decoder
\item Attention loss for decoder
\end{itemize}

Like other end-to-end speech recognizers, Conformer does not need an
explicit language model or a finite state transducer to guide its speech
inference. 
Its encoder directly infers from the speech a sequence of acoustic latent
representations, $\mathbf{Y}  = \{ \mathbf{y}_t | \mathbf{y}_t = \mathop{\mbox{Enc}}(\mathbf{x_t}), t = 1 \dots T \}$, by feeding the
Mel-frequency filter bank coefficients $\mathbf{X}  = \{\mathbf{x}_t| t=1 \dots
T\}$ with $T$~frames into it. 
Note that although Conformer shrinks to $1/4$ in its
convolution block, to simplify the presentation, we do not express this in the formula.  

To predict the next word~$\hat{w}$, the decoder uses cross attentions via
$\mathbf{Y}  = \{\mathbf{y_t}  | t = 1
\dots T\}$ conditioned on history words $W^h$ as 
\begin{equation}
  \hat{w} = \mathop{\arg\max} \limits_{w \in \mathcal{V}} \mathop{\mbox{Att}}(\mathbf{Y} | W^h).
  \label{eq_crossatt}
\end{equation}

We then apply the CTS mask $M_{\mathit{CTS}}$ to the acoustic encoding sequence~$\mathbf{Y}$ as
\begin{equation}
  \hat{w} = \mathop{\arg\max} \limits_{w \in \mathcal{V}} \mathop{\mbox{Att}}(M_{\mathit{CTS}} \cdot \mathbf{Y} | W^h).
  \label{eq_cts_crossatt}
\end{equation}

The components are unchanged from the original Conformer except 
for the CTS function itself and the cross-attention parameters in the decoder,
as described in the following sections.

\subsection{CTS Masking}

\begin{figure*}[bht]
  \centering
  \includegraphics[width=\linewidth]{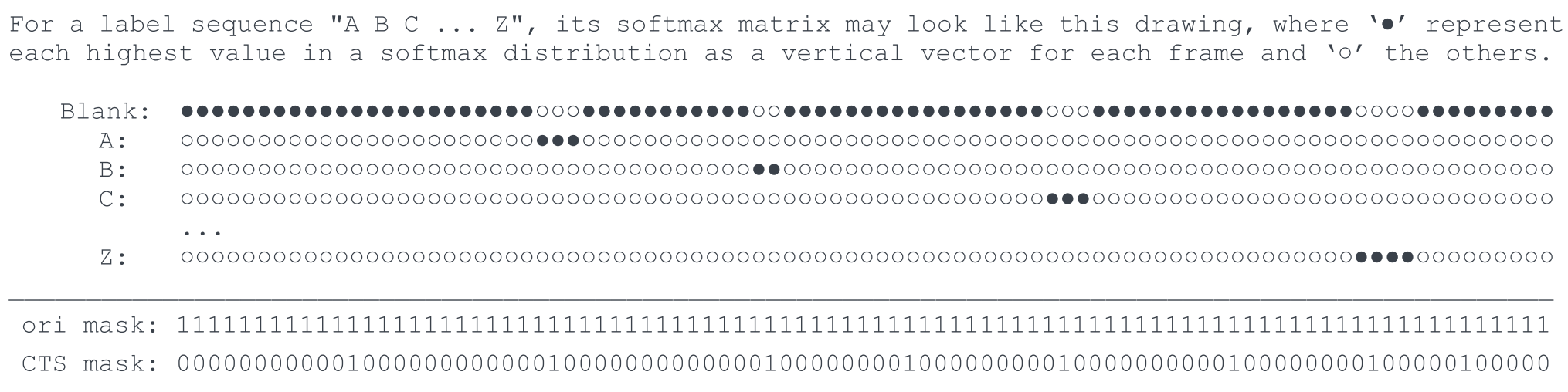}
  \caption{Effect of CTS masking}
  \label{fig_cts_mask}
\end{figure*}

The purpose of CTS masking is to use the mask function $M_{\mathit{CTS}}$ to
change the latent sequence~$\mathbf{Y}$ from the frame basis to the 
segment basis, a sequence of boolean values of length~$T$.   
Each boolean value in $M_{\mathit{CTS}}$ indicates whether the latent representation
of the frame's encoder should be used to compute cross-attention softmax values in the
decoder. 
Frames with a negative boolean value are not used in the decoder to
predict the next word, 
reducing the computational effort of the decoder. 
We expect its affect to accuracy can be resolved; 
in any case, this can be corrected after fine-tuning.    
Figure~\ref{fig_cts_mask} illustrates the concept of CTS masking.

The CTS mask $M_{\mathit{CTS}}$ is generated according to the following algorithm. 
Firstly, compute softmax distributions for each frame $\mathbf{y}_t$ in the acoustic encoding sequence
with the help of the linear transformation matrix $H_{CTC}$ in the CTC-loss component.
It map $y_t$ from the dimension of latent representation vector for each frame, to the size of word-piece labels in vocabulary $\mathcal{V}$, like as in Eq.~(\ref{eq_softmax}):
\begin{equation}
  p_{w,t} = \mathop{\mbox{softmax}}\limits_{w \in \mathcal{V}}(H_{CTC} \cdot \mathbf{y_t}).
  \label{eq_softmax}
\end{equation}
Then, find out the belonging label which has maximal softmax score for each frame by Eq.~(\ref{eq_argmax}):
\begin{equation}
  \hat{w}_t = \mathop{\arg\max}\limits_{w \in \mathcal{V}}(p_{w,t}).
  \label{eq_argmax}
\end{equation}
Then, do the following steps: 
\begin{itemize}
\item {\bf{Segmentation}}: for any continuing frames with the same maximum label ``$\hat{w}_{t_s} = \cdots = \hat{w}_{t_e}$'', aggregate them as one segment. 
\item {\bf{Representation}}: for each segment, find the frame $\hat{t}$ with the maximal softmax score by Eq.~(\ref{eq_repre})
	 and select its vector $y_{\hat{t}}$ to represent the whole segment.
\item {Masking}: mark frames that are not selected with a negative mask value 
     and skip these during decoder training.   	 
\end{itemize}
\begin{equation}
  \hat{t} = \mathop{\arg\max}\limits_{{t_s} \leq t \leq {t_s}}(p_{\hat{w},t}).
  \label{eq_repre}
\end{equation}
Note that CTS masking resembles the
max-pooling effect over time for its segmental
selection and representation processing 
as illustrated in Eq.~\ref{eq_maxpool}:
\begin{equation}
  \mathbf{Y'} = 
  \mathop{\mbox{CTS}}(\mathbf{Y})
  \cong
  \{ \mathbf{y'}_i | \mathop{\mbox{maxpool}}\limits_{{t_{s_i}} \leq t \leq {t_{s_i}}}(\mathbf{y}_t), i=1 \dots S \} , 
\label{eq_maxpool}
\end{equation}
where $i$ is the segment index. 

After augmenting Conformer training with CTS masking, 
most encoding frames are skipped among blank segments.
This introduces some influences to its decoder performance.
To smooth this change, 
we propose changes to the CTS-Conformer training, as described below. 

\subsection{CTS-Conformer Training}

CTS-Conformer attaches a CTS mask to the original Conformer. 
However, this mask requires a well-trained CTC element 
to segment and select positive mask values 
over the frames that are representative of the whole acoustic sequence. 
Therefore, we train the Conformer with the CTC component, 
attach the CTS mask, and then refine the CTS-Conformer parameters.

Again, the attachment of the CTS mask
changes the input to cross attention of the decoder,  
leading to increased attention loss. 
If still using the hybrid losses for training 
the whole CTS-Conformer parameters, 
it may introduce errors to affect the well-trained CTC encoder's parameters. 
Hence, 
in the beginning of attaching the CTS mask,
the fine-tuning can be applied to the CTS-Conformer with frozen encoder parameters. 
After training well of decoder's cross-attention parameters, 
it may be stable to release encoder's parameters for 
fully hybrid-loss training. 
At this time, 
  there is little change to the encoder parameters,   
and the CTC loss is also more stable.
We call this way a `two-step fine-tuning' approach.
It was proven useful in our experiments for CTS-Conformer's training. We thus adopted the following training procedure
for CTS-Conformer parameter training 
with alternative fine-tuning approaches.
\begin{enumerate}
\item Train standard Conformer model 
\item Add CTS mask
\item Fine-tune encoder and decoder parameters. 
This can be done in three different ways:
\begin{itemize}
\item {\bf `Full fine-tuning'}: fine-tune the normal way for the whole set of parameters.
\item {\bf `Fine-tuning with frozen encoder'} (or {\bf `free-enc'} in short): 
  fine-tune the decoder parameters only.    
\item {\bf `Two-step fine-tuning'}: first fine-tune with the frozen encoder and then do full fine-tuning.
\end{itemize}
\end{enumerate}

Due to the nature of the CTS algorithm, the effect of fine-tuning 
depends heavily on the CTC accuracy.
Without low CTC loss, segmentation and representation selection
perform poorly, which can influence the CTS mask and subsequent 
cross attentions 
  that depend on the mask.   
The effect of fine-tuning can also depend on the size of the dataset used for fine-tuning. 
Many ASR applications have limited training resources, such as in spoken
language understanding (SLU). 
With such small fine-tuning datasets,
encoder adaptation may have little effect on the parameters, 
thus yielding poor fine-tuning results.
We conducted experiments on two datasets in order to vberify CTS-Conformer performance broadly:
LibriSpeech, with around one thousand hours of speech, 
and the FluentSpeech speech data for SLU research. 
FluentSpeech includes less than fifteen hours of speech for fine-tuning.   
For FluentSpeech ASR, we additionally use LibriSpeech to pretrain the Conformer.
Below we discuss the difference in their fine-tuning effects.

\section{Experiments}

\subsection{Datasets}

For our experiments, we used the open-source LibriSpeech ASR corpus 
\footnote{``http://www.openslr.org/12''}~\cite{Panayotov2015}, as the major speech dataset with which to 
train the Conformer. 
This is a corpus of read speech derived from audiobooks, comprising approximately 1000~hours
of 16\,kHz read English speech. 
For training, we use the 960-hour training set of clean and noisy speech for training, 
the 5.4-hour `dev-clean' set for validation, and the 5.4-hour `test-clean' set and 5.3-hour `test-other' set
for testing. 

We also used the
Fluent Speech Commands dataset \footnote{``https://fluent.ai/research/fluent-speech-commands/''},
or FluentSpeech, as the other data example in our experiments. 
This is a corpus
of SLU data for spoken commands to smart homes or virtual assistants,
for instance, ``put on the music'' or ``turn up the heat in the kitchen''.
Each audio utterance is labeled with three slots: \emph{action}, \emph{object}, and \emph{location}. 
A slot takes one of multiple values: for instance, \emph{location} can
take the values ``none'', ``kitchen'', ``bedroom'', or ``washroom''. 
There are few training speech data,
and all are spoken based on a very limited vocabulary and sentence expressions. 
We took the well-trained Conformer (by LibriSpeech) and fine-tuned it with
FluentSpeech training data, and then used the resultant Conformer to train our
CTS-Conformer models to evaluate our approach on this small dataset. 

\begin{table}[thb]
  \caption{Datasets}
  \label{tab:datasets}
  \centering
  \begin{tabular}{lrrr}
    \toprule
    \textbf{Dataset} & \textbf{Spkrs} & \textbf{Utts} & \textbf{Hours}                \\
    \midrule
    \textbf{LibriSpeech} & &  &     \\
    \hspace{5mm} train      & 5,466  & 281,241     & 960.7       \\
    \hspace{5mm}  dev-clean & 97    & 2,703     & 5.4       \\
    \hspace{5mm}  test-clean & 87    & 2,620     & 5.4       \\
    \hspace{5mm}  test-other & 90    & 2,939     & 5.3       \\
    \textbf{FluentSpeech} & &  &     \\
    \hspace{5mm} train  & 77    & 23,132    & 14.7      \\
    \hspace{5mm} valid  & 10    & 3,118     & 1.9       \\
    \hspace{5mm} test   & 10    & 3,789     & 2.4       \\
    \bottomrule
  \end{tabular}
\end{table}

\subsection{CTS-Conformer Fine-tuning}

We started with Conformer training. 
And after adding the CTS mask to the well-trained Conformer, 
we fine-tuned it to optimize its accuracy. 
Both models were built with the ESPNet 
toolkit \footnote{``https://github.com/espnet/espnet''} using Mel-frequency filter bank
(Fbank) feature vectors as input:
this is a sequence of 83-dimensional feature vectors 
with a 25-ms window size and 10-ms window shifts. 
The model used 2-layer convolutional neural networks (CNN) as the
frontend, each of which had 256 filters with 3$\times$3 kernel size
and 2$\times$2 stride; thus the time reduction of the frontend was 1$/$4.
Our vocabulary was a list of 5000 byte-pair-encoded~\cite{shibata1999byte}
wordpiece labels.
The above models both contained 12~layers of self-attention encoder blocks and 6~layers
of decoder blocks, each of which contained a self-attention layer
and an encoder-decoder cross-attention layer. 
The multi-head attentions in both models had
four heads, 256 attention dimensions, and 1024-dimensional feedforward networking.
They were trained using hybrid CTC/attention loss with a CTC weight of~0.3.

First, we used the LibriSpeech `test-clean' and `test-other' sets to compare the different models; the 
results are presented in Table~\ref{tab:asr_librispeech}.
The best CTS-Conformer results are obtained by the two-step fine-tuning approach.
Virtually no loss of accuracy for CTS-Conformer fine-tuned with the method, except on `test-clean' WER was reduced from 3.63\% to 3.73\% as decoding at a beamwidth of 10.  
No accuracy loss at all for `test-other' set. 
Larger improvements were found for both test sets at a beamwith of 1, relative word error rate (WER) reductions of 10\% and 6\% respectively were seen on `test-clean` and `test-other' sets.

\begin{table}[ht]
  \caption{ASR WER (\%) on LibriSpeech with different models using 
various beamwidths}
  \label{tab:asr_librispeech}
  \centering
  \begin{tabular}{lccc}
    \toprule
    \multicolumn{1}{r}{Beamwidth}  & 1 & 4 & 10 \\
    \midrule
    & \multicolumn{3}{c}{`test-clean' set}  
    \\
    \cline{2-4}
    \textbf{CTS-Conformer} & & & \\
    \hspace{0.2cm} w/ full fine-tuning 	& 4.41 & 3.98 & 3.78 \\
    \hspace{0.2cm} w/ frozen-enc fine-tuning & 4.74 & 4.08 & 3.98 \\
    \hspace{0.2cm} w/ 2-step fine-tuning 	& \textbf{4.27} & \textbf{3.77} & 3.73 \\
    \textbf{Conformer}  				  & 4.75 & 3.78 & \textbf{3.63} \\
    \midrule
    & \multicolumn{3}{c}{`test-other' set} 
    \\
    \cline{2-4}
    \textbf{CTS-Conformer} & & & \\
    \hspace{0.2cm} w/ full fine-tuning 	& 9.55 & \textbf{8.85} & 8.75 \\
    \hspace{0.2cm} w/ frozen-enc fine-tuning & 9.93 & 8.99 & 8.96 \\
    \hspace{0.2cm} w/ 2-step fine-tuning 	& \textbf{9.47} & \textbf{8.85} & \textbf{8.74} \\
    \textbf{Conformer}  				  & 10.08 & 9.11 & 8.79 \\
    \bottomrule
  \end{tabular}
\end{table}
\begin{table}[ht]
  \caption{Same comparison as in Table~\ref{tab:asr_librispeech} but on FluentSpeech}
  \label{tab:asr_fluentspeech}
  \centering
  \begin{tabular}{lccc}
    \toprule
    \multicolumn{1}{r}{Beamwidth}  	& 1 & 4 & 10 \\
    \midrule
    \textbf{CTS-Conformer} 	& & & \\
    \hspace{0.5cm} w/ full fine-tuning 	& 1.11 & 1.09 & 1.10 \\
    \hspace{0.5cm} w/ frozen-enc fine-tuning & 1.12 & 1.08 & 1.09 \\
	\hspace{0.5cm} w/ 2-step fine-tuning 	&  1.11 & 1.09 & 1.09 \\
    \textbf{Conformer}   	& 1.09 & 1.09 & 1.11 \\
    \bottomrule
  \end{tabular}
\end{table}

We then performed the same comparison on
FluentSpeech, as shown in Table~\ref{tab:asr_fluentspeech}.
Note that FluentSpeech's Conformer was derived from the Conformer 
listed in Table~\ref{tab:asr_librispeech}, which was pre-trained with
LibriSpeech and then fine-tuned with FluentSpeech.  
CTS-Conformer was trained with the same steps as above, 
i.e., fine-tuning CTS-Conformer model based on the Conformer parameters well-trained already.  
Its all WERs are almost small, 
likely because the FluentSpeech corpus is relatively simple,
with its small vocabulary and limited expressions.
Only small gain and loss of 0.02\% absolute values were seen in the table as comparing WERs of two-step fine-tuned CTS-Conformer with those of Conformer.

To further understand the effect of
CTS masking on cross attention, we used the CTS mask on the Conformer model without
fine-tuning and indeed noted a significant change in accuracy.
Table~\ref{tab:cts_investigate} presents the results: 
clearly, in addition to the additional substitution errors, 
there were also far more insertion errors. 
Such results demonstrate that the refinement of parameters is thus important after changing the input to the decoder. 
 
\begin{table}[htb]
  \caption{ASR results by attaching the CTS mask to Conformer but without fine-tuning of parameters, 
  compared with the one decoding with no CTS masking. Beamwidth = 4.}
  \label{tab:cts_investigate}
  \centering
  \begin{tabular}{crrrr}
    \toprule
    Data        &    \multicolumn{2}{c}{LibriSpeech} &  \multicolumn{2}{c}{FluentSpeech} \\
    Model       & \textbf{Conf.} & +\textbf{CTS} & \textbf{Conf.} & +\textbf{CTS} \\
    \midrule
    WER (\%)    & 3.78 & 61.12  & 1.09 & 3.64 \\
    sub         & 2.62 & 13.27  & 0.38 & 0.39 \\
    del         & 0.76 & 4.00  & 0.07 & 0.11 \\
    ins         & 0.41 & 43.58  & 0.64 & 3.14 \\
    \bottomrule
  \end{tabular}
\end{table}


\subsection{Decoding Time}

To the important evaluation of decoding time consumption, 
we evaluated the decoding real-time factor (RTF) of CTS-Conformer against
Conformer with CPU decoding. The time-saving percentage of using CTS masking is
listed in Table~\ref{tab:computation}. 

\begin{table}[htb]
  \caption{Real-time factor (RTF) of CPU decoding of LibriSpeech and FluentSpeech
with Conformer and CTS-Conformer on an
Intel Xeon\textregistered{} Gold 6130 CPU @ 2.10GHz}
  \label{tab:computation}
  \centering
  \begin{tabular}{lrrr}
    \toprule
    Decoding & \textbf{Conf.} & \textbf{CTS-Conf.} & Time savings \\
    \midrule
    LibriSpeech & & & \\
	\hspace{5mm} Beam 1 & 0.183 & 0.165 & 10\% \\
	\hspace{5mm} Beam 4 & 0.610 & 0.487 & 20\% \\
	\hspace{5mm} Beam 10 & 1.460 & 1.212 & 16\% \\
    \midrule
    FluentSpeec & & & \\
	\hspace{5mm} Beam 1 & 0.130 & 0.135 & -4\% \\
	\hspace{5mm} Beam 4 & 0.387 & 0.346 & 11\% \\
	\hspace{5mm} Beam 10 & 0.860 & 0.729 & 15\% \\
    \bottomrule
  \end{tabular}
\end{table}

All RTFs for LibriSpeech were reduced largely. 
A reduction of 20\% was seen at a beamwidth of 4. 
It is important because we usually use this beamwidth in decoding of large vocabulary continuous ASR applications. 
A comparison of FluentSpeech decoding times between Conformer and CTS-Conformer
reveals instabilities when the beamwidth is 1. 
An increase from 0.130 to 0.135 was found. 
Although CTS masking shortens the length of the encoding sequence, 
it does not always lead to reduced decoding times.
When CTS-Conformer's fine-tuning results in an encoder with higher CTC loss, it introduces increasing paths from CTC decoding to be rescored by attention decoding. 
Thus, it may cause longer computational consumption at some time.
Especially, FluentSpeech is a small dataset with a limited vocabulary,  
there are many other words in the vocabulary could not be fine-tuned.
Such an imbalance could cause problems like insertion errors in decoding, 
resulting in incorrectly increased decoding times.
When decoding with a higher beamwidth than 1, 
such a phenomenon is smoothed out and disappears.

\section{Conclusions}

We add CTS masking to the Conformer model to construct CTS-Conformer models. 
This improves the Conformer decoder's efficiency by
simplifying the cross attention's input acoustic sequence. We also propose
several fine-tuning approaches. In our experiments, when using attention-rescoring
decoding with beamwidth of four, CTS-Conformer reduces
the overall computations on LibriSpeech and FluentSpeech by 20\% or 10\%  
respectively without harming ASR accuracy.

\bibliographystyle{IEEEtran}

\end{document}